\documentclass[aps,prl, preprintnumbers,amsmath,amssymb,latexsym,array,enumerate,letter,twocolumn,superscriptaddress]{revtex4-2}

\usepackage{amssymb}
\usepackage{amsmath}
\usepackage{epsfig}
\usepackage{hyperref}
\usepackage{breakurl}
\usepackage{xcolor}

\usepackage{graphicx, subfigure, placeins, float}

\usepackage{extarrows}

\usepackage{multirow} % Required for multirows
\usepackage{longtable} % and/or

\graphicspath{{./figure/}}

\newcommand{\itbf}[1]{\textbf{\textit{#1}}}

\begin{document}

\title{Three-loop color-kinematics duality: A 24-dimensional solution space \\ induced by new generalized gauge transformations}
\author{Guanda Lin}
\email{linguandak@pku.edu.cn}
\affiliation{School of Physics, Peking University, Beijing 100871, China}
\affiliation{CAS Key Laboratory of Theoretical Physics, Institute of Theoretical Physics, Chinese Academy of Sciences,  Beijing, 100190, China}
\author{Gang Yang}
\email{yangg@itp.ac.cn}
\affiliation{CAS Key Laboratory of Theoretical Physics, Institute of Theoretical Physics, Chinese Academy of Sciences,  Beijing, 100190, China}
\affiliation{School of Physical Sciences, University of Chinese Academy of Sciences, Beijing 100049, China}
\affiliation{School of Fundamental Physics and Mathematical Sciences, Hangzhou Institute for Advanced Study, UCAS, Hangzhou 310024, China}
\affiliation{International Centre for Theoretical Physics Asia-Pacific, Beijing/Hangzhou, China}
\author{Siyuan Zhang\vspace{2mm}}
\email{zhangsiyuan@itp.ac.cn}
\affiliation{CAS Key Laboratory of Theoretical Physics, Institute of Theoretical Physics, Chinese Academy of Sciences,  Beijing, 100190, China}
\affiliation{School of Physical Sciences, University of Chinese Academy of Sciences, Beijing 100049, China}

%%%%%%%%%%%%%%%%%%%%%
\begin{abstract}

We obtain full-color three-loop three-point form factors of the stress-tensor supermultiplet and also of a length-3 half-BPS operator in ${\cal N}=4$ SYM based on the color-kinematics duality and on-shell unitarity. 
The integrand results are verified by all planar and non-planar unitarity cuts, and they satisfy the minimal power-counting of loop momenta and diagrammatic symmetries. 
Interestingly, these three-loop solutions, while manifesting all dual Jacobi relations, contain a large number of free parameters; in particular, there are 24 free parameters for the form factor of stress-tensor supermultiplet. 
Such degrees of freedom are due to a new type of generalized gauge transformation associated with the operator insertion for form factors.
We also perform numerical integration and obtain consistent full-color infrared divergences and the known planar remainder.
The form factors we obtain can be understood as the ${\cal N}=4$ SYM counterparts of three-loop Higgs plus three-gluon amplitudes in QCD and are expected to provide the maximally transcendental parts of the latter.

\end{abstract}

\maketitle

%%%%%%%%%%%%%%%%%%%%
\section{Introduction}
\label{sec:introduction}

\noindent
Remarkable progress in our understanding of fundamental interactions has been made in both theoretical and experimental aspects in the last decades. One central impetus for these developments is the uncovering of surprising and intriguing mathematical structures hidden in microscopic scattering processes, where  the maximally supersymmetric Yang-Mills theory ($\mathcal{N}=4$ SYM) has been an ideal laboratory. 
In the planar limit, the amplitudes are relatively well-understood and even certain all-loop integrand constructions \cite{ArkaniHamed:2010kv, Arkani-Hamed:2016byb} and non-perturbative functional results \cite{Basso:2013vsa, Basso:2015uxa} have been achieved, 
thanks to the underlying integrability of planar $\mathcal{N}=4$ SYM, see \cite{Beisert:2010jr} for a review.
However, while going beyond the planar limit, the inclusion of color degrees of freedom complicates the problem, because the latter breaks many of the planar symmetries. As a result, understanding the non-planar sector of the theory remains a very challenging task.

Taking advantage of the color ``complications",
a remarkable duality between color and kinematics was discovered by Bern, Carrasco and Johansson \cite{Bern:2008qj, Bern:2010ue}. 
The color-kinematics (CK) duality proposes that the gauge theory amplitudes can be organized in terms of trivalent graphs such that
the kinematic numerators satisfy identities in one-to-one correspondence with color Jacobi identities. When combined together with the generalized unitarity method \cite{Bern:1994zx, Bern:1994cg, Britto:2004nc}, this duality makes possible 
high-loop constructions of gauge amplitudes with full-color dependence, see \emph{e.g.~}high-loop amplitudes in SYM \cite{Carrasco:2011mn, Bern:2012uf, Bern:2013qca, Bern:2014sna, Johansson:2017bfl, Bern:2017ucb, Kalin:2018thp} and pure YM \cite{Boels:2013bi, Bern:2013yya, Bern:2015ooa, Mogull:2015adi}, and also Sudakov form factors up to five loops in ${\cal N}=4$ SYM \cite{Boels:2012ew,Yang:2016ear, Lin:2020dyj}. Apart from the significance to gauge theories, the duality also builds a bridge connecting gauge and gravity theories: the gravity amplitudes can be directly constructed from the Yang-Mills amplitudes in the CK-dual representation, via the ``double copy'' \cite{Bern:2010ue, Bern:2010yg}. The so-called ``double copy" property has many impressive utilities, for example, for understanding the ultraviolet properties of gravity theories \cite{Bern:2012uf,Bern:2012cd,Bern:2012gh,Bern:2013uka,Bern:2017yxu,Bern:2017ucb,Bern:2018jmv,Herrmann:2018dja}. A recent and extensive review of the duality and its applications can be found in \cite{Bern:2019prr}.
 
The CK duality has been proved at tree-level using string or gauge theory methods \cite{BjerrumBohr:2009rd, Stieberger:2009hq, Feng:2010my}. 
However, at loop level the duality is still a conjecture and has only been shown by explicit constructions.
Thus, it is very important to explore more examples and see to what extent the duality applies.
It is worth pointing out that a loop representation fully manifesting CK duality is generally non-trivial to reach. For example, it has proven difficult to find such a representation for the five-loop four-point amplitude in ${\cal N}=4$ SYM \cite{Bern:2017yxu, Bern:2017ucb}. Another example is the all-plus two-loop five-gluon amplitudes in pure YM theory: numerators with 12 powers of loop momenta that are much more than that of Feynman diagrams have to be used to realize the duality \cite{Mogull:2015adi}. 

In this paper, we obtain new three-loop solutions which manifest the color-kinematics duality for a class of three-point form factors in ${\cal N}=4$ SYM. Interestingly, the results contain a large number of free ``gauge" parameters; for example, for the form factor of the stress-tensor supermultiplet, there are 24 free parameters.
We would like to emphasize that our results belong to the ``simplest" type of solutions, in the sense that: they maintain all diagrammatic symmetries, and they satisfy the minimal power-counting behavior expected in ${\cal N}=4$ SYM. 
As we will discuss later, these free parameters originate from a new type of generalized gauge transformation, which is induced by the operator insertion in form factors.

Concretely, the form factors considered here describe the interaction between three on-shell states $\Phi_i$ and a gauge invariant operator ${\cal O}$ (see \cite{Yang:2019vag} for an introduction):
\begin{equation}
    \mathcal{F}_{\mathcal{O}}(1, 2, 3 ; q) = \int d^{D} x e^{-i q \cdot x}\langle \Phi_1 \Phi_2 \Phi_3|\mathcal{O}(x)| 0\rangle \,.
\end{equation}
Here we consider half-BPS operators
 $\mathcal{O}_{L}={\rm tr}(\phi^L)$, with $L=2$ and $3$; in particular, ${\cal O}_2$ is a component of the stress-tensor supermultiplet.
Besides the theoretical significance to CK duality, these form factors also bear phenomenological interest, due to their close relation 
to the Higgs-plus-three-gluon amplitudes in QCD. In particular, both these two $\mathcal{N}=4$ form factors were found to coincide with
the maximally transcendental part (\emph{i.e.}~functionally the most complicated part) of the corresponding QCD results up to two-loop order
\cite{Brandhuber:2012vm, Gehrmann:2011aa, Brandhuber:2014ica, Brandhuber:2017bkg,  Jin:2018fak},
providing examples of the general principle of maximal transcendentality \cite{Kotikov:2002ab, Kotikov:2004er}.
As for higher-loop orders, the planar three-point form factor of $\mathcal{O}_2$
has been computed
via bootstrap recently up to five loops \cite{Dixon:2020bbt} using the input from the form factor operator product expansion (OPE) \cite{Sever:2020jjx, Sever:2021nsq}. 
Our result provides for the first time the three-loop non-planar correction, 
for which powerful methods such as the OPE bootstrap are not yet applicable.
For the form factor of $\mathcal{O}_3$, the two-loop result was given in \cite{Brandhuber:2014ica}, and here we provide the new three-loop result.

Below we first give a brief review of CK duality and introduce our computational strategy. Then we explain our construction of CK-dual solutions.
We further perform numerical integration, check the full-color infrared (IR) subtraction and obtain finite remainders.
Finally we discuss the interpretation of free parameters. The complete CK-dual solutions are given in the ancillary files.

%%%%%%%%%%%%%%%%%%%%%%%%%%%%%%%%%%%
%%%%%%%%%%%%%%%%%%%%%%%%%%%%%%%%%%%
\section{Review and strategy}
\label{sec:review}

%%%%%%%%%%%%%%%%%%%%%%%%%%%%%%%%%%%%%%%%%%
\begin{figure}[t]
\centering
\includegraphics[clip,scale=0.5]{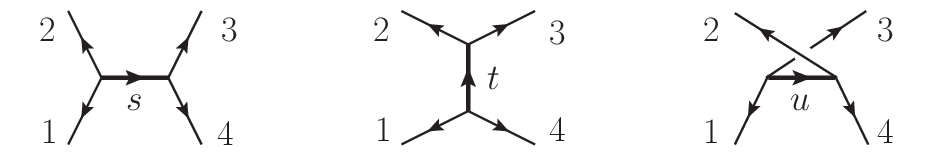}
\caption[a]{Trivalent graphs for four-point tree amplitudes.}
\label{fig:tree4CK}
\end{figure}
%%%%%%%%%%%%%%%%%%%%%%%%%%%%%%%%%%%%%%%%%%

\noindent
An instructive example to illustrate the color-kinematics duality is the four-gluon tree amplitude. It is possible to represent the amplitude in terms of three cubic graphs shown in Fig.~\ref{fig:tree4CK}:
\begin{equation}
    \itbf{A}^{(0)}_{4}=\frac{C_s N_s}{s}+\frac{C_t N_t}{t}+\frac{C_u N_u}{u}\,,
\end{equation}
where $C_i$ are color factors given as products of structure constants $\tilde f^{abc}$ corresponding to each trivalent vertex,
and $N_i$ are kinematic numerators that contain intrinsic physical information. 
Here we use the normalization ${\rm tr}(T^a T^b)= \delta^{ab}$ and $\tilde f^{abc} = {\rm tr}(T^a [T^b, T^c])$.
The CK duality requires that the numerators should satisfy the dual Jacobi relations parallel to the corresponding color Jacobi relations as \cite{Bern:2008qj}
\begin{equation} 
\label{eq:dualJacobi}
C_s = C_t + C_u  \ \  \Rightarrow \ \  N_s = N_t + N_u \,.
\end{equation}

%%%%%%%%%%%%%% FIGURE  %%%%%%%%%%%%%%%%%%%%%%%
\begin{figure}[t]
\centering
\includegraphics[clip,scale=0.33]{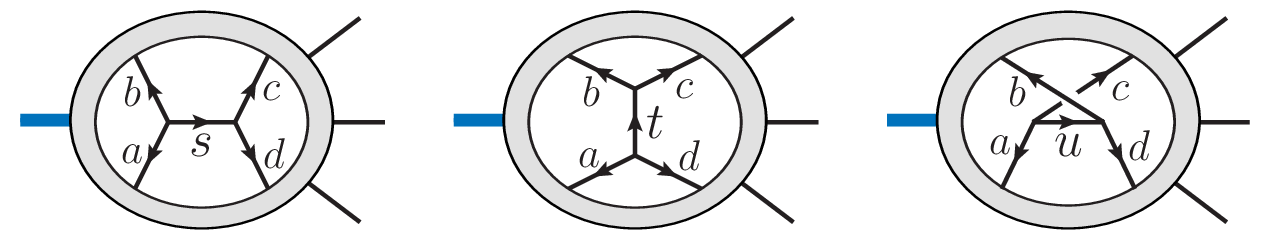}
\caption[a]{Loop diagrams related by Jacobi relation.}
\label{fig:loopCK}
\end{figure}
%%%%%%%%%%%%%%%%%%%%%%%%%%%%%%%%%%%%%%%%%%

While the CK duality has been proved at tree level, the striking point is that the duality can be generalized to loop level.
Consider trivalent loop diagrams for form factors shown in Fig.~\ref{fig:loopCK}: 
every internal propagator (not directly connected to the $q$-leg) is associated to a four-point tree sub-graph.
The three diagrams in Fig.~\ref{fig:loopCK} containing $s, t, u$-channel sub-graphs are related by a color Jacobi relation,
and CK duality requires that their numerators satisfy the dual Jacobi relation as:
\begin{equation}
\label{eq:dualJacobiLoop}
N_s(l) = N_t(l) + N_u(l) \,.
\end{equation}
Note that the four-point sub-diagrams in Fig.~\ref{fig:loopCK} have four \emph{off-shell} legs $l_i, i ={a,b,c,d}$, 
thus it is a highly non-trivial generalization from \eqref{eq:dualJacobi} to \eqref{eq:dualJacobiLoop}.
These dual Jacobi relations will play a central role in the following three-loop constructions. 

The general strategy of our constructions involves two major steps. 
The first step is to construct an ansatz of the loop integrand corresponding to a set of cubic graphs. By imposing the dual Jacobi relations \eqref{eq:dualJacobiLoop}, the numerators of different topologies are interlocked with each other, and an ansatz for the CK-dual integrand can be obtained efficiently. 
In the second step, we solve the ansatz by imposing physical constraints, where the main tool is the generalized unitarity method \cite{Bern:1994zx, Bern:1994cg, Britto:2004nc}.
Below we implement the above strategy to construct the three-loop form factor solutions.
Readers are also referred to \cite{Bern:2012uf, Carrasco:2015iwa, Yang:2019vag} for further details of general constructions.

%%%%%%%%%%%%%%%%%%%%%%%%%%%%%%%%%%%
\section{Constructing CK-dual solutions}
\label{sec:results}

\begin{figure}[t]
\centering
\includegraphics[clip,scale=0.36]{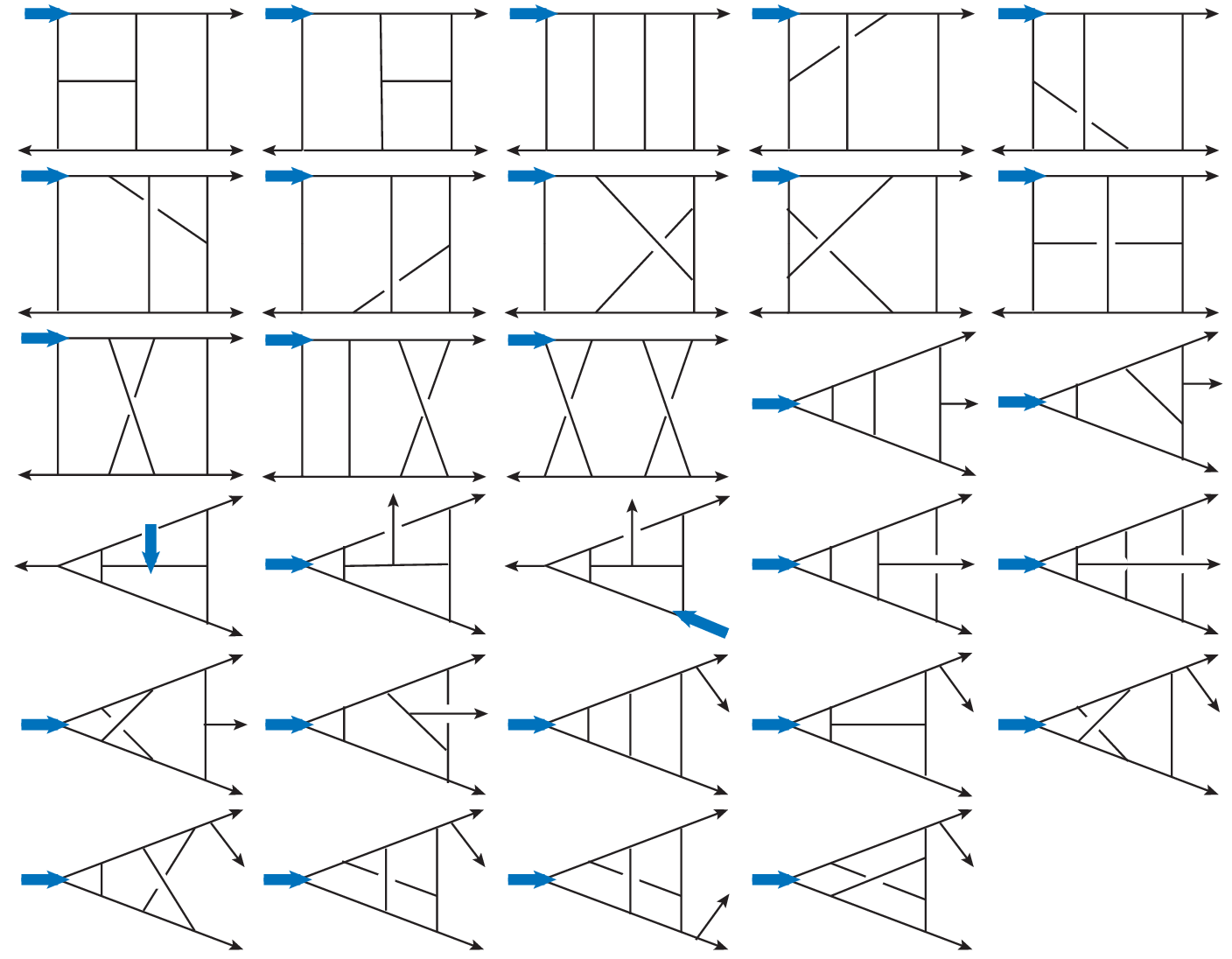}
\caption{Trivalent topologies for the form factor of  $\operatorname{tr}(\phi^2)$. }
\label{fig:phi2tops}
\end{figure}

\noindent
We firstly discuss the form factor of ${\rm tr}(\phi^2)$. 
The starting point is to construct a set of trivalent graphs for the three-loop integrand. 
Each diagram contains four external legs: three on-shell legs $p_i$ and one off-shell leg $q$ associated to the operator.
Following the empirical experience from the known high-loop CK-dual solutions \cite{Bern:2012uf, Boels:2012ew,Yang:2016ear}, we exclude graphs with tadpole, bubble and triangle sub-graphs, unless the triangle is attached to the $q$-leg. 
We find that there are 29 trivalent topologies to consider, as shown in Fig.~\ref{fig:phi2tops}.

\begin{figure}[b]
\centering
\includegraphics[clip,scale=0.48]{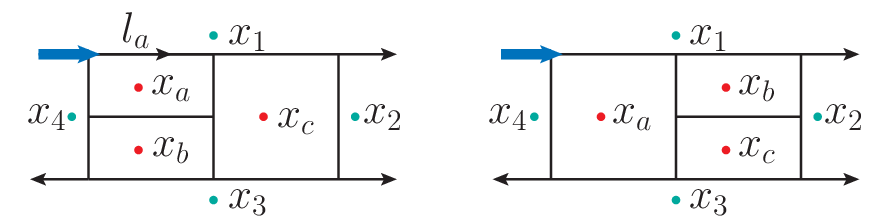}
\caption{Master graphs for $\operatorname{tr}(\phi^2)$ form factor. }
\label{fig:phi2master}
\end{figure}

Next we consider Jacobi relations for all four-point sub-graphs of these topologies. It turns out that all topologies
can be generated by two planar topologies shown in Fig.~\ref{fig:phi2master}, which are called master graphs. 
Once knowing the numerators of these two master graphs, all other numerators can be deduced via dual Jacobi relations \eqref{eq:dualJacobiLoop}.

%%%% All Cubic Graphs L=2 %%%%%
%%%%%%%%%%%%%%%%%%%%%%%%%%%%%%%%%
%%%%%%%%%%%%%%%%%%%%%%%%%%%%%%%%%

%%%% Master Graphs L=2 %%%%%
%%%%%%%%%%%%%%%%%%%%%%%
%%%%%%%%%%%%%%%%%%%%%%%

To construct ansatz for two planar master numerators, it is convenient to use zone variables $x_i$ \cite{Drummond:2006rz}, such as $x_{a1}^2 = (x_a - x_{1})^2=l^2_a$ shown in  Fig.~\ref{fig:phi2master}.
Based on the nice UV properties of ${\cal N}=4$ SYM and half-BPS operators, we impose the minimal power-counting condition for loop momenta: a one-loop $n $-point sub-graph carries no more than $n-4 $ powers of the corresponding loop momentum \cite{Bern:2012uf}, with an exception that if the sub-graph is a one-loop form factor, the maximal power is $n-3$ \cite{Boels:2012ew}. 
Specifically, 
for $N_1$ (the first master in Fig.~\ref{fig:phi2master}), $x_a,x_c $ can appear at most once, so $(x^2_{ac})^1$ or  $(x^2_{ai})^1 (x^2_{ci})^1$, with $i=1,2,3,4$, are allowed; whereas terms containing $x_b $ or containing more than one $x_a$ or $x_c $, such as $(x^2_{ac})^2,(x^2_{a1})^2 $, are forbidden. 
For the other master numerator $N_2$, only $x_a$ can appear with maximal power 2, thus only $(x^2_{ai})^n$ with power $n=2,1,0 $ can appear. 
With these constraints, we obtain an ansatz form as  linear combinations of monomials of zone variables with an overall  dimension $[x^8]$, and two master numerators have 201 and 115 free parameters respectively.
Then we get an ansatz of the complete integrand with 316 parameters.

\begin{figure}[t]
\centering
\includegraphics[clip,scale=0.45]{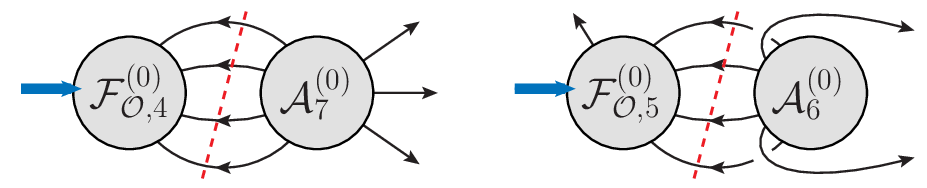}
\caption{Quadruple cuts for 3-loop 3-point form factors.}
\label{fig:quadcuts}
\end{figure}

Given the ansatz, we further apply various constraints to fix the parameters. 
We first require numerators to respect the automorphism symmetries of corresponding graphs, see \emph{e.g.}~\cite{Bern:2012uf, Boels:2012ew}.
This provides substantial constraints on the ansatz and reduces the number of parameters to 105. 
To  fix the remaining parameters, we apply (generalized) unitarity cuts \cite{Bern:1994zx, Bern:1994cg, Britto:2004nc}, where two examples of the most complicated quadruple cuts are illustrated in Fig.~\ref{fig:quadcuts}.
Interestingly, after imposing a spanning set of unitarity cuts, there are still 24 parameters left. 
We point out that our integrand correctly reproduce not only planar but also non-planar cuts, 
which ensures that the unitarity constraints are complete. 
Finally, we check that all dual Jacobi relations are satisfied.
Thus we get the physical three-loop integrand solution with $24$ free parameters that also manifests CK-duality.

%%%% most complex cuts %%%%%
%%%%%%%%%%%%%%%%%%%%%%%%%%%%%
%%%%%%%%%%%%%%%%%%%%%%%%%%%%%

%%%% All Cubic Graphs L=3 %%%%%
%%%%%%%%%%%%%%%%%%%%%%%%%%%%%%%%%
\begin{figure}[b]
\centering
\includegraphics[clip,scale=0.275]{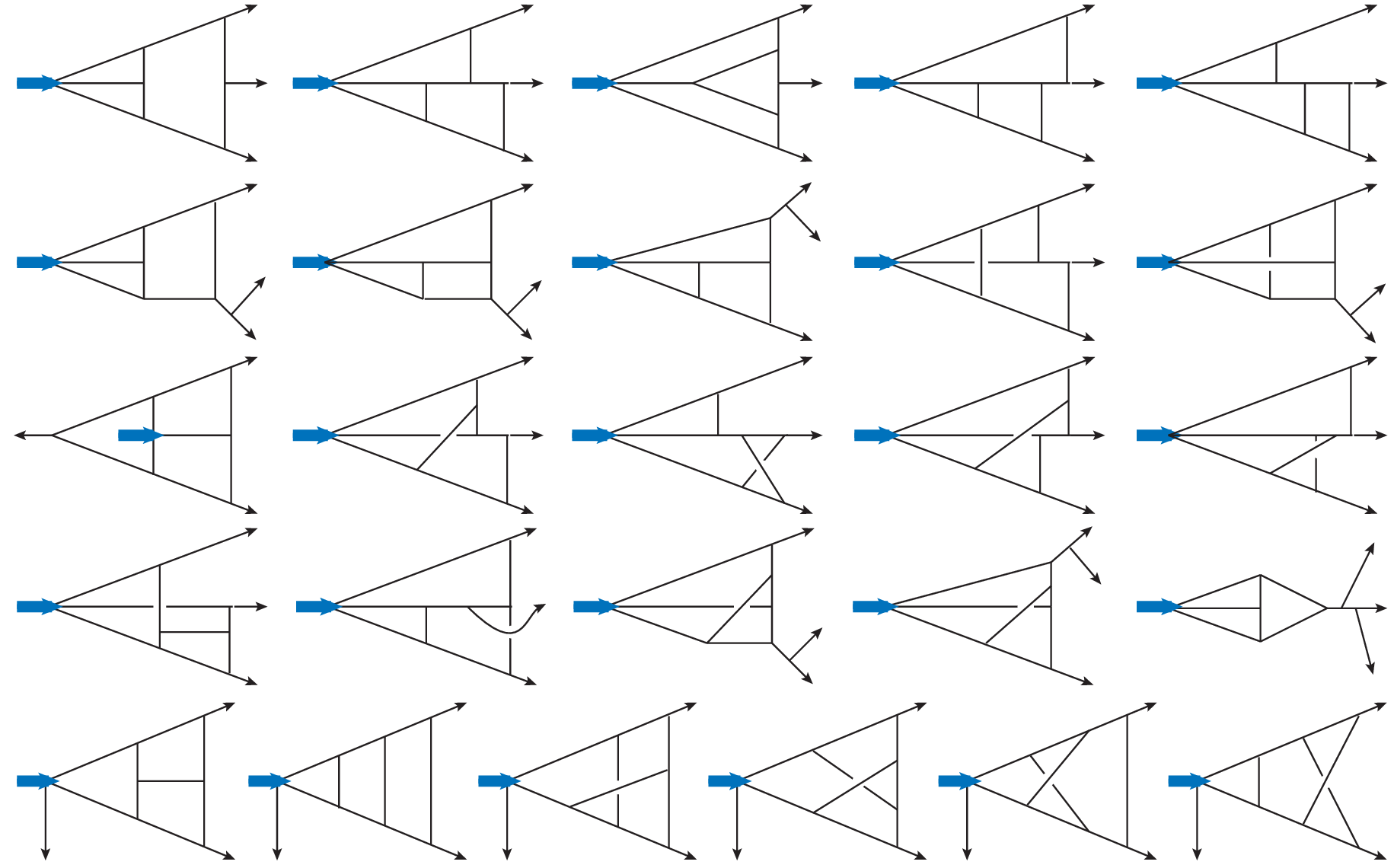}
\caption{Trivalent topologies for the form factor of  $\operatorname{tr}(\phi^3)$. }
\label{fig:phi3tops}
\end{figure}
%%%%%%%%%%%%%%%%%%%%%%%%%%%%%%%%%

%%%% Master Graphs L=3 %%%%%
%%%%%%%%%%%%%%%%%%%%%%%
\begin{figure}[t]
\centering
\includegraphics[clip,scale=0.31]{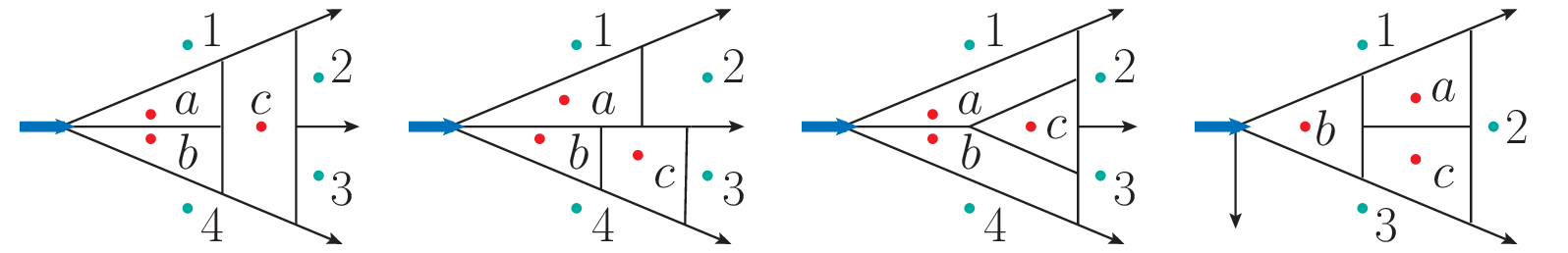}
\caption{Master graphs for $\operatorname{tr}(\phi^3)$ form factor.}
\label{fig:phi3master}
\end{figure}
%%%%%%%%%%%%%%%%%%%%%%%

Following the above procedure, we also construct the three-loop integrand for the form factor of ${\rm tr}(\phi^3)$. 
Since the operator contains three scalars, in the trivalent diagrams the $q$-leg must be associated with a four-point vertex, as shown in  Fig.~\ref{fig:phi3tops}.
In this case, one can divide the three-loop topologies into two classes. 
The first class consists of integrals that involve all three external on-shell legs in the interactions, as shown in the first four rows in Fig.~\ref{fig:phi3tops}. 
In the second class, depicted by the diagrams in the last row of Fig.~\ref{fig:phi3tops}, one of the external on-shell legs is connected directly to the $q$-leg; these diagrams correspond to single-scale integrals. 
The dual Jacobi relations are decoupled between these two classes;
consequently, the master topologies should be selected separately, as presented in Fig.~\ref{fig:phi3master}: the first three planar masters are for the first class, and the last one for the second class. 

Since the procedure of constructing and solving the ansatz is similar to the ${\rm tr}(\phi^2)$ case, we will not go into 
details and only give a brief summary for the ${\rm tr}(\phi^3)$ case. 
An ansatz satisfying minimal power-counting for the four master numerators has in total 295 parameters. After imposing the symmetry and unitarity constraints, we get the CK-dual integrand solution with 10 free parameters.

For both form factors, we give solutions of master numerators in the ancillary files.

%%%%%%%%%%%%%%%%%%%%%
\section{Results and checks}
\label{sec:solutionandcheck}

\noindent
The final CK-dual integrands of three-loop form factors can be summarized as:
\begin{equation}  
\label{eq:fullcolor-3loop}
 \itbf{F}_{\mathcal{O}_{L}, 3}^{(3)} = {\cal F}_{\mathcal{O}_{L}, 3}^{(0)}   \sum_{\sigma_3} \sum_{i} \int \prod_{j=1}^{3} d^D \ell_j {1\over S_i} \,\sigma_3 \cdot {C_i \, N_i \over \prod_{\alpha_i} P^2_{\alpha_i}} \,,
\end{equation}
where $S_i$ are symmetry factors that remove the overcounting from the automorphism symmetries of the graphs, and $\sigma_3$ are the permutations of external on-shell momenta $p_i, i=1,2,3$ \footnote{The permutation operator $\sigma_3$ acts on $N_i, P^2_{\alpha_i} $ and $C_i $ and changes external momenta and color indices associated to them respectively. For the form factor of ${\rm tr}(\phi^2) $, acting $\sigma_3$ on $N_i $ may give an extra sign, \emph{i.e.}~$\sigma_3\cdot N_i(p_1,p_2,p_3)=\text{sign}(\sigma_3)N_i(\sigma_3(p_1),\sigma_3(p_2),\sigma_3(p_3))$, since the tree level form factor (factorized as $\mathcal{F}^{(0)}$ gives a $\text{sign}(\sigma_3)$ when permuting external lines). }.
Explicit expressions of the symmetry factors $S_i $, color factors $C_i $, propagator lists $P_{\alpha_i} $ and numerators $N_i $ are given in the ancillary files. 

The physical form factors constructed via the unitarity method are expected to be independent of the remaining free parameters.
It turns out that all parameters cancel simply at the integrand level, and such a cancellation is related to the generalized gauge transformations induced by the operator insertion.
We will come back to this in the discussion section.
The simplified integrands of form factors in $N_c$ expansion can be given as:
\begin{align}  
\itbf{F}_{\mathcal{O}_2, 3}^{(3)} & = {\cal F}_{\mathcal{O}_2, 3}^{(0)} {\tilde f}^{a_1a_2a_3} \big( N_c^3 {\mathcal{I}}^{(3)}_{\mathcal{O}_2} + 12 N_c {\mathcal{I}}_{\mathcal{O}_2,\rm NP}^{(3)} \big) \,, \\
\itbf{F}_{\mathcal{O}_3, 3}^{(3)} & = {\cal F}_{\mathcal{O}_3, 3}^{(0)} {\tilde d}^{a_1a_2a_3} N_c^3 {\mathcal{I}}^{(3)}_{\mathcal{O}_3} \,,
\end{align} 
where ${\tilde d}^{a_1a_2a_3}= {\rm tr}(T^{a_1}T^{a_2}T^{a_3}) +{\rm tr}(T^{a_1}T^{a_3}T^{a_2})$. 

We see that the form factor of ${\rm tr}(\phi^2)$ contains a non-planar three-loop correction.
Notably, non-dipole corrections of IR structures appear for the first time at three loops \cite{Almelid:2015jia}. 
To compare with these structures, we take a numerical approach to calculate the contributed three-loop integrals, using packages FIESTA \cite{Smirnov:2015mct} and pySecDec \cite{Borowka:2017idc}. The evaluation of the non-planar parts of these integrals turns out to be highly involved. To overcome this difficulty, 
we managed to organize the integrand into uniformly transcendental integrals to the necessary extent, which significantly improves the efficiency of computation; such a simplification has also been observed in the four-loop Sudakov form factor computation \cite{Boels:2017skl, Boels:2017ftb}. 
Our results give consistent IR divergences, for both the planar \cite{Bern:2005iz} and non-planar parts \cite{Almelid:2015jia, Henn:2016jdu}. Moreover, 
the three-loop planar remainder for the form factor of ${\rm tr}(\phi^2)$ confirms the recent bootstrap computation \cite{Dixon:2020bbt} (using also data from \cite{Spradlin:2008uu, Gehrmann:2011xn}). 
All these provide strong consistency checks of our results. Some details of the numerical checks are provided in the supplemental material.
More details on the simplification of the integrand and numerical computations will be given in \cite{Lin:2021qol}.

%%%%%%%%%%%%%%%%%%%%%
\section{Discussion}
\label{sec:discussion}

\noindent
An interesting finding of this work is that the integrand solutions contain a large number of free parameters, while at the same time manifesting all dual Jacobi relations. 
Practically, this is a very appealing property for the high-loop construction using the CK-dual ansatz.
Such free parameters can be understood as deformations of the integrand that do not change the final form factor result. 
Below we discuss the origin of these deformations.

We first point out that the free parameters we find have \emph{no} relation to the traditional gauge transformations that correspond to changing external polarization vectors as $\varepsilon^\mu_i \rightarrow \varepsilon^\mu_i + \alpha p^\mu_i$ for arbitrary $\alpha$.
This is simply because the loop corrections are independent of polarization vectors.

Another type of integrand transformation is the so-called  \emph{generalized gauge transformation} (GGT) \cite{Bern:2010ue}. 
For example, one can deform the numerators associated with $s,t,u$-channel trivalent graphs as
\begin{equation}
\label{eq:generalGT}
N_{s}\rightarrow N_s + s \Delta,\quad  N_{t}\rightarrow N_t - t \Delta,\quad  N_{u}\rightarrow N_u - u \Delta\,,
\end{equation}
for arbitrary $\Delta$ without changing the amplitude or form factor results.
At loop level, such a transformation typically breaks the dual Jacobi relations, since $s+t+u\neq0$, see Fig.~\ref{fig:loopCK}.

%%%%%%%%%%%%%%%%%%%%%%%%%%%%
\begin{figure}[b]
\centering
\includegraphics[clip,scale=0.35]{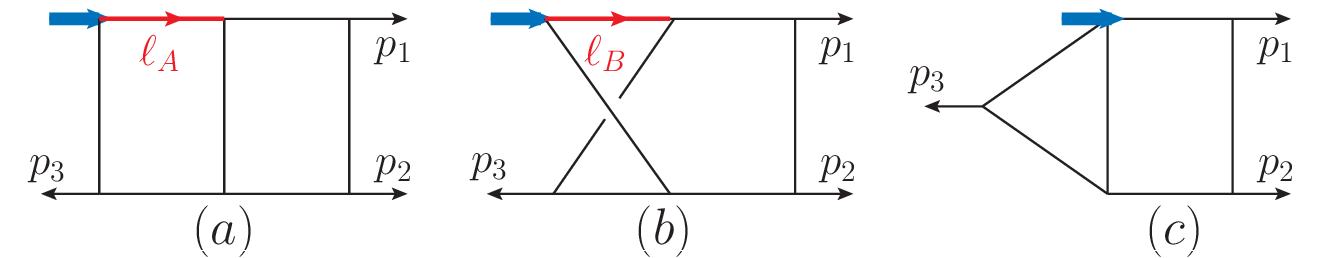}
\caption{Graph examples of 2-loop form factor of ${\rm tr}(\phi^2)$. Fig.~(a) and (b) have the same color factor, and after shrinking the propagators labeled by $\ell_A$ and $\ell_B$, they both reduce to Fig.~(c). }
\label{fig:GTff}
\end{figure}
%%%%%%%%%%%%%%%%%%%%%%%%%%%%

For form factors, a new type of generalized gauge transformations appears, due to the insertion of local operators. 
To illustrate this point, let us consider the simpler two-loop three-point form factor of ${\rm tr}(\phi^2)$. In this case, we find that the two-loop CK-dual representation (with minimal power-counting) also contains 4 free parameters. Consider the graphs in Fig.~\ref{fig:GTff}: the first two topologies share the same color factor 
\begin{equation}
\label{eq:operatorinduced}
C_{\rm a} = C_{\rm b} \,,
\end{equation}
since the color factor of the local operator ${\rm tr}(\phi^2)$ is a $\delta$-function in color space. 
This implies that one can make a deformation of the corresponding numerators as:
\begin{equation}
\label{eq:generalGT2}
N_{\rm a}\rightarrow N_{\rm a} + \ell_A^2 \Delta,\quad  N_{\rm b}\rightarrow N_{\rm b} - \ell_B^2 \Delta, 
\end{equation}
which leaves the full integrand unchanged:
\begin{equation}
C_{\rm a}\, I_{\rm a}[\ell_A^2\Delta] - C_{\rm b}\, I_{\rm b}[\ell_B^2\Delta] = (C_{\rm a} - C_{\rm b})\, I_{\rm c}[\Delta]=0\,.
\end{equation}
Here $I_{\rm{a,b,c}}$ are integrals related to the topologies in Fig.~\ref{fig:GTff}. 
Note that \eqref{eq:generalGT2} has different color-algebraic origin from \eqref{eq:generalGT}.
To distinguish these two transformations, we call \eqref{eq:generalGT2} the \emph{operator-induced} generalized gauge transformation (GGT), while referring \eqref{eq:generalGT} as \emph{Jacobi-induced} GGT.

The high-dimensional solution space in the CK-dual representation is closely related to the operator-induced GGT for form factors. 
Indeed, in order to show that the integrands in the solution space (with different choices of the free parameters) are equivalent, one must apply the color relations \eqref{eq:operatorinduced} and the operator-induced GGT for the considered form factors.
 Explicit two-loop examples are given in the supplemental material.

Finally, we recall that gravity amplitudes can be obtained through double copy of gauge amplitudes in the CK-dual representation, where the Jacobi-induced GGT leads to diffeomorphism invariance in gravity \cite{Bern:2019prr}.
Given the CK-dual solutions for form factors, one may ask if there is a physical meaning of performing
 double copy in this case.
In order to have a consistent double-copy result, 
where color factors $C_i$ are replaced by kinematic numerators ${\tilde N}_i$:
\begin{equation}
\sum_i {C_i N_i \over \prod D_{i, \alpha}}  \ \ \Rightarrow \ \ \sum_i {{\tilde N}_i N_i \over \prod D_{i, \alpha}} \,,
\end{equation}
the ${\tilde N}_i$ should satisfy all corresponding color relations including those operator-induced color relations such as in the case of Fig.~\ref{fig:GTff}: 
\begin{equation}
C_{\rm a} = C_{\rm b} \ \ \Rightarrow \ \ {\tilde N}_{\rm a} = {\tilde N}_{\rm b}  \,.
\end{equation}
Otherwise, the free parameters in $N_i$ will not cancel in the double-copy result.
We find that for the two-and-three-loop three-point form factors, there is no numerator solution satisfying such a requirement.
This implies that the operator-induced GGT may not give diffeomorphism invariance in gravity through double copy, 
which seems consistent with the known fact that local diffeomorphism-invariant operators do not exist in gravity.
It would be interesting to explore this point further, since
the argument here does not exclude the possibilities that the double-copy picture may apply to form factors in special choices of gauge, as well as their possible connections to certain non-local observables in gravity.

\vskip .2 cm 

%%%%%%%%%%%%%%%%%%%%%%%%%%%%%%%%%%%%%%%%%
{\it Acknowledgments.}
It is a pleasure to thank Yuchen Ding, Yuanhong Guo, Song He and Yanqing Ma for interesting discussions.
This work is supported in part by the National Natural Science Foundation of China (Grants No.~11822508, 11935013, 12047503), 
and by the Key Research Program of the Chinese Academy of Sciences, Grant NO. XDPB15.
We also thank the support of the HPC Cluster of ITP-CAS and CAS Xiandao-1 computing environment.


%%%%%%%%%%%%%%%%%%%%%%%%%%%%%%%%%%%%%%%%%%%
%\onecolumngrid

\appendix

\newpage
%%%%%%%%%%%%%%%%%%%%%%%%%%%%%%%%%%%
%%%%%%%%%%%%%%%%%%%%%%%%%%%%%%%%%%%
\onecolumngrid
\section{Supplemental material}

\noindent
In this supplemental material, we first discuss the numerical integration of the results and check the IR structure and planar remainder,  then we discuss the free-parameter cancellations in detail in the two-loop form factor of ${\rm tr}(\phi^2)$.

\subsection{A. Integration and Checks}

\noindent
The loop correction functions $\mathcal{I}^{(3)}$ are given in Eq.~(6) and (7) in the paper.
We compute integrals numerically at a special kinematic point $s_{12}=s_{23}=s_{13}=-2$, using packages FIESTA \cite{Smirnov:2015mct} and pySecDec \cite{Borowka:2017idc}.  
Both the $N_c$-leading results of $\mathcal{O}_{2,3}$ and the $N_c$-subleading result for $\mathcal{O}_{2}$ ($\mathcal{O}_{3}$ has no $N_c$-subleading contribution) are calculated up to $\epsilon^0$, shown in Table~\ref{tab:numres} and Table~\ref{tab:numres_np}, respectively.
We point out that the computation of non-planar integrations are quite involved, taking about $10^7$ core hours to get the current precision, even with the improvement by using uniformly transcendental representation. 

Given the numerical data, we compare them with the universal IR structures and the known planar finite remainder. We begin with the planar part. The planar IR divergent part can be captured by the BDS ansatz \cite{Bern:2005iz}:
\begin{align}\label{eq:bds}
{\mathcal{I}}^{(3)}(\epsilon)
= -\frac{1}{3}\big({\mathcal{I}}^{(1)}(\epsilon)\big)^{3} + {\mathcal{I}}^{(2)}(\epsilon) {\mathcal{I}}^{(1)}(\epsilon)+f^{(3)}(\epsilon) {\mathcal{I}}^{(1)}(3 \epsilon)
+ \mathcal{R}^{(3)}+\mathcal{C}^{(3)}+O(\epsilon)\,,
\end{align}
where
\begin{align}
\label{eq:f3}
f^{(3)}(\epsilon)=4\left(\frac{11}{2}\zeta_4 +(6 \zeta_5+5\zeta_2\zeta_3)\epsilon+(c_1\zeta_6+c_2\zeta_3^2)\epsilon^2\right), 
\end{align}
with  (see  \cite{Spradlin:2008uu})
\begin{equation}
\label{eq:c1c2num}
(c_1\zeta_6+c_2\zeta_3^2) = 85.263 \pm 0.004\,,
\end{equation}
and $\mathcal{C}^{(3)}$ is a finite constant. The divergent part given by \eqref{eq:bds} depends only on lower-loop results \cite{Brandhuber:2012vm} and is ready to be calculated. We find that the divergent parts of our results match with the BDS predictions perfectly.

For $\mathcal{O}_2$, we can also compare the finite order result with \cite{Dixon:2020bbt}, where $\mathcal{R}_3^{(3)}$ is calculated via the OPE bootstrap method. 
In order to do so, one needs to determine the constant $\mathcal{C}_3^{(3)}$ in \eqref{eq:bds},
which is defined such that it is independent of the number of external legs and the corresponding remainder satisfies $\mathcal{R}_n^{(3)} \rightarrow \mathcal{R}_{n-1}^{(3)}$ in the collinear limit. 
It can be fixed by using the three-loop Sudakov form factor result  (where $n=2$) \cite{Gehrmann:2011xn}. Imposing that the Sudakov remainder is zero in \eqref{eq:bds}, one has
\begin{equation}
{\cal R}_{\mathcal{O}_2,2}^{(3)} = 0 \quad \Rightarrow \quad \mathcal{C}^{(3)}_{\mathcal{O}_2} = -38.252 \pm 0.004 \,.
\end{equation}
Thus all the elements in $\eqref{eq:bds}$ are clarified and one can consider the $\epsilon^{0}$ order. For example, at kinematic point $s_{12}=s_{23}=s_{13}$, using the remainder result from bootstrap in  \cite{Dixon:2020bbt}, one expects that (the error is due to \eqref{eq:c1c2num} as well as small numerical errors of high-$\epsilon$ order of lower-loop input in BDS ansatz)
\begin{equation}
{\mathcal{I}}^{(3),{\rm bootstrap}}_{\mathcal{O}_2} \big|_{\epsilon^0} = -336.716 \pm 0.013 \,,
\end{equation}
which is consistent with our numerical computation 
\begin{equation}
{\mathcal{I}}^{(3),{\rm num}}_{\mathcal{O}_2} \big|_{\epsilon^0} = -336.51\pm 0.59 \,.
\end{equation}

% %%%%%%%%%%%%%%%
 \begin{table*}[t]
     \centering
     \caption{Numerical results for planar three-point form factors.     \label{tab:numres}} 
     \vskip .1 cm
%   %
     \begin{tabular}{|c|c|c|c|c|c|c|c|}
     \hline \hline
 Form factor &
 \multicolumn{7}{c|}{$\mathcal{I}^{(3)}_{\operatorname{tr}(\phi^2)}$} \\\hline
       $(s_{12},s_{23},s_{13})$  & $\epsilon^{-6}$ & $\epsilon^{-5}$ & $\epsilon^{-4}$ & $\epsilon^{-3}$ & $\epsilon^{-2}$ & $\epsilon^{-1}$ & $\epsilon^{0}$ \\\hline
       (-2,-2,-2)  & $-4.5 $ & $9.357488 $ & $-22.61361 $ & $55.8891 $ & $-77.252 $ & $92.943 $ & $-336.51 $ \\\hline
       est. error & $8.0\times10^{-10} $ & $2.3\times10^{-5} $ & $ 3.3\times 10^{-4} $ & $0.0021 $ & $0.012 $ & $0.078 $ & $0.59 $ \\\hline\hline
 Form factor &
 \multicolumn{7}{c|}{$\mathcal{I}^{(3)}_{\operatorname{tr}(\phi^3)}$} \\\hline
       $(s_{12},s_{23},s_{13})$  & $\epsilon^{-6}$ & $\epsilon^{-5}$ & $\epsilon^{-4}$ & $\epsilon^{-3}$ & $\epsilon^{-2}$ & $\epsilon^{-1}$ & $\epsilon^{0}$ \\\hline
       (-2,-2,-2)  & -4.5 & $9.357488$ & -6.02807 & 31.5028 & 19.5617 & 123.565 & 217.11 \\\hline
      est. error & $2\times 10^{-7}$ & $9.4\times 10^{-6}$ & $8.1\times 10^{-5}$ & $5.4\times 10^{-4}$ & 0.0035 & 0.023 & 0.21\\\hline
     \end{tabular}

 \end{table*}
% %%%%%%%%%%%%%%%

% %%%%%%%%%%%%%%%
 \begin{table*}[t]
    \centering
    \caption{Numerical results for non-planar three-point form factor of stress-tensor supermultiplet.     \label{tab:numres_np}} 
    \vskip .1 cm
%   %
     \begin{tabular}{|c|c|c|c|c|c|c|c|}
    \hline \hline
 Form factor &
 \multicolumn{7}{c|}{$\mathcal{I}^{(3)}_{\operatorname{tr}(\phi^2),{\rm NP}}$} \\\hline
       $(s_{12},s_{23},s_{13})$  & $\epsilon^{-6}$ & $\epsilon^{-5}$ & $\epsilon^{-4}$ & $\epsilon^{-3}$ & $\epsilon^{-2}$ & $\epsilon^{-1}$ & $\epsilon^{0}$ \\\hline
       (-2,-2,-2)  & $-2.3\times 10^{-7}$ & $5.8\times10^{-6}$ & $3.8\times10^{-5}$ & $5.6\times10^{-4}$ & $-0.0010$ & $-9.99$ & $-265.3$ \\\hline
      est. error & $1.2\times 10^{-6}$ & $2.4\times 10^{-5}$ & $3.0\times 10^{-4}$ & $2.5\times10^{-3}$ & $0.020$ & $0.18$ & $1.8$\\\hline
     \end{tabular}
 \end{table*}
% %%%%%%%%%%%%%%%

Next we consider non-planar IR divergences. The full-color three-loop IR divergence can be factorized out as 
\begin{equation}
\label{eq:fullcolorir}
    \itbf{F}(p_i,a_i,\epsilon)={\itbf{Z}}(p_i,\epsilon)\itbf{F}^{\rm fin}(p_i,a_i,\epsilon)\,,
\end{equation}
where ${\itbf{Z}}$ takes the form \cite{Almelid:2015jia} (see also \cite{Henn:2016jdu})
\begin{equation}\label{eq:zfactor}
    \itbf{Z}=\mathcal{P} \exp\left[\sum_{\ell=1}^{\infty}g^{2\ell}\left(\text{dipole terms}+\frac{1}{\ell \epsilon} \mathbf{\Delta}^{(\ell)}\right)\right]\,,
\end{equation}
with $g^2 = \frac{g_{\rm YM}^2 }{(4\pi)^2}(4\pi e^{- \gamma_{\text{E}}})^\epsilon$.
The dipole terms can be completely fixed by cusp and collinear anomalous dimensions and contribute only to the $N_c$-leading three-point form factors. The $\mathbf{\Delta}$ terms represent the non-dipole contributions starting from three-loop order \cite{Almelid:2015jia} and contribute only to the $N_c$-subleading form factors.
Moreover, for the three-point form factor, because of the small number of external lines, one only needs to consider
\begin{equation}
\label{eq:3loopdelta}
    \mathbf{\Delta}_{3}^{(3)}= \alpha 
    \sum_{\substack{(i,j,k)\in (1,2,3)\\j<k}} {f}_{abe} {f}_{cde} (\mathbf{T}^{a}_i\mathbf{T}^{d}_i+\mathbf{T}^{d}_i\mathbf{T}^{a}_i)\mathbf{T}^{b}_{j}\mathbf{T}^{c}_{k}\,,
\end{equation}
where $\alpha = - 8 (\zeta_5+2\zeta_2\zeta_3)$ and the color operators  act as $\mathbf{T}_{1}^{a} T^{a_{1}}=- i f^{a a_{1} c} T^{c}$. 
Consequently, the non-planar correction $\mathcal{I}^{(3)}_{\mathcal{O}_2,\mathrm{NP}}$ should have only divergences at $\epsilon^{-1}$  order with coefficient 
\begin{equation}\label{eq:npirepm1}
    \mathcal{I}^{(3), {\rm analytic}}_{\mathcal{O}_2,\mathrm{NP}}\Big|_{\epsilon^{-1}}= -2 (\zeta_5+2\zeta_2\zeta_3) =-9.983\,,
\end{equation}
which also lies in the error range of our result:
\begin{equation}\label{eq:npirepm1-num}
\mathcal{I}^{(3), {\rm num}}_{\mathcal{O}_2,\mathrm{NP}}\Big|_{\epsilon^{-1}}= - 9.99 \pm 0.18\,.
\end{equation}
We would like to  point out further the contributing integrals cancel delicately to get the correct IR structure. Take the $\epsilon^{-1}$ order as an example: there are dozens of (non-planar) integrals to consider, each of which takes a value of several $100 \sim 1000$ at the $\epsilon^{-1}$ order; a delicate cancellation is required to reproduce the small number in \eqref{eq:npirepm1-num}.

%%%%%%%%%%%%%%%%%%%%%%%%%%
%%%%%%%%%%%%%%%%%%%%%%%%%%

\subsection{B. CK-dual two-loop three-point form factor of $\operatorname{tr}(\phi^2)$: free parameters and cancellation} % (fold)
\label{sub:two_loop_three_point_form_factor_of_phi2_with_four_ck_preserving_parameters}

\noindent
As mentioned in the discussion section, a careful revisit of the construction for the two-loop three-point form factor of $\operatorname{tr}(\phi^2) $ shows that the generic CK-dual solution contains 4 parameters. 
Similar to the three-loop form factors discussed in the paper, this two-loop solution maintains all diagrammatic symmetries and satisfies the minimal power-counting behavior expected in ${\cal N}=4$ SYM.

The complete CK-dual solution can be given as
\begin{equation}  
\label{eq:fullcolor-2loop}
 \itbf{F}_{\mathcal{O}_{2}, 3}^{(2)} = {\cal F}_{\mathcal{O}_{2}, 3}^{(0)}   \sum_{\sigma_3} \sum_{i=1}^6 \int \prod_{j=1}^{3} d^D \ell_j {1\over S_i} \,\sigma_3 \cdot {C_i \, N_i \over \prod_{\alpha_i} P^2_{\alpha_i}} \,,
\end{equation}
The six topologies and various factors are explicitly given in Table~\ref{tab:resphi2-2loop}, where $\Gamma_{i,123},N_{i,123},C_{i,123} $ and $S_i $ denote topologies, numerators,
color factors and symmetry factors respectively. 
The subscription $123 $ labels the ordering of external lines and its permutations may give an extra sign, see the footnote [49] in the paper. As for the notation: in the numerators, $s_{ij}=2p_i\cdot p_j, \tau_{mi}=2l_m\cdot p_i$, with $i,j=1,2,3 $ and $m=a,b $, and $q=p_1+p_2+p_3$.

%%%%%%%%%%%%%%%%%%%  
\begin{longtable}{|c|c|c|c|}
    \caption{CK-dual representation of the two-loop three-point form factor of $\operatorname{tr}(\phi^2)$.}
    \label{tab:resphi2-2loop}
    \endfirsthead
    \endhead
    \hline \begin{tabular}{cc}   \\ ~  \end{tabular} $\Gamma_{i,123} $ &  $N_{i,123} $ & $C_{i,123} $ & $S_i$  \\
    \hline \\
%%%%%%%%%%%%%%% top 1 information %%%%%%%%%%%%%%%
    \begin{tabular}{cc}    
        \includegraphics[width=0.20\linewidth]{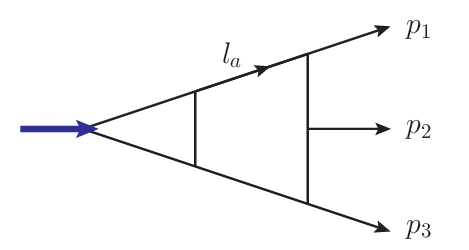}  \\
        (1)  
    \end{tabular}
    &
    $\begin{aligned}
          & q^2 s_{12} s_{23}/2 \\
        - & c_1 
        \begin{aligned}[t]
            \bigl(
              & (s_{23}(s_{12}+s_{13})-2s_{12}s_{13})) (l_a-p_1-p_2)^2 \\
            + & (s_{12}(s_{13}+s_{23})-2s_{13}s_{23}) (l_a-p_1)^2 \\
            + & s_{12} (s_{13}-s_{23}) (l_a-q)^2 + s_{23} (s_{13}-s_{12}) l_a^2 
            \bigr)
        \end{aligned} \\
        - & c_2 
        \begin{aligned}[t]
            \bigl(
              & (2s_{23}^2-s_{12}^2-s_{13}^2)(l_a-p_1-p_2)^2 \\
            + & (2s_{12}^2-s_{13}^2-s_{23}^2) (l_a-p_1)^2 \\
            + & (s_{13}^2-s_{23}^2) (l_a-q)^2 + (s_{13}^2-s_{12}^2) l_a^2 
            \bigr)
        \end{aligned}
    \end{aligned} $   
    & $2N_c^2 \tilde f^{a_1 a_2 a_3} $ & 2  \\
    \hline \\
%%%%%%%%%%%%%%% top 2 information %%%%%%%%%%%%%%%
    \begin{tabular}{cc}    
        \includegraphics[width=0.20\linewidth]{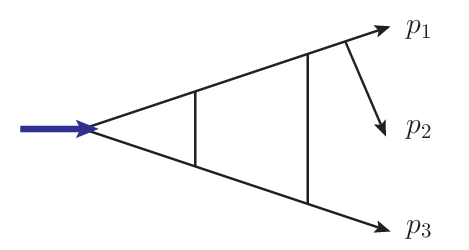}  \\
        (2)  
    \end{tabular}
    &
    $\begin{aligned}
          & q^2 s_{12} (s_{13}+s_{23})/2 \\
        + & c_1 
        \begin{aligned}[t]
            (s_{12}(s_{13}+s_{23}) - 2s_{13}s_{23}) s_{12}
        \end{aligned} \\
        + & c_2
        \begin{aligned}[t]
            (2 s_{12}^2-s_{13}^2-s_{23}^2) s_{12}
        \end{aligned}
    \end{aligned} $
    & $4N_c^2 \tilde f^{a_1 a_2 a_3} $ & 2 \\
    \hline \\
    %%%%%%%%%%%%%%% top 3 information %%%%%%%%%%%%%%%
    \begin{tabular}{cc}    
        \includegraphics[width=0.20\linewidth]{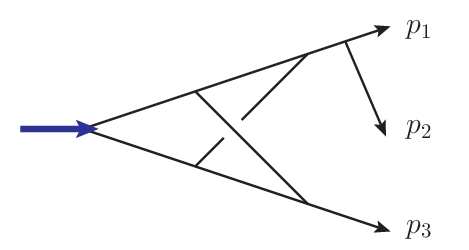}  \\
        (3)  
    \end{tabular} 
    &
    $\begin{aligned}
          & q^2 s_{12} (s_{13}+s_{23})/2 \\
        + & c_1 
        \begin{aligned}[t]
            (s_{12} (s_{13}+s_{23})-2 s_{13} s_{23}) s_{12}
        \end{aligned} \\
        + & c_2 
        \begin{aligned}[t]
            (2 s_{12}^2-s_{13}^2-s_{23}^2) s_{12}
        \end{aligned}
    \end{aligned} $
    & $2N_c^2 \tilde f^{a_1 a_2 a_3} $ & 4 \\
    \hline \\
    %%%%%%%%%%%%%%% top 4 information %%%%%%%%%%%%%%%
    \begin{tabular}{cc}    
        \includegraphics[width=0.20\linewidth]{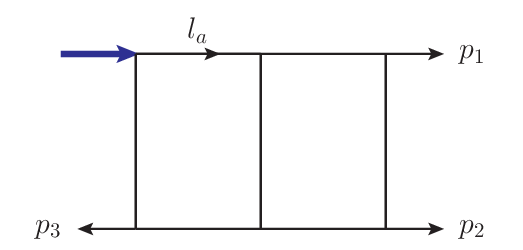}  \\
        (4)  
    \end{tabular} 
    & 
    $\begin{aligned}
          & s_{12}(s_{23}\tau_{a2}-s_{13}\tau_{a1})/2 \\
        + & c_1
        \begin{aligned}[t]
            s_{12} (s_{13}-s_{23}) \bigl((l_a-p_1-p_2)^2 - (l_a-q)^2 \bigr)
        \end{aligned} \\
        + & c_2
        \begin{aligned}[t]
            (s_{13}^2-s_{23}^2) \bigl((l_a-p_1-p_2)^2 - (l_a-q)^2 \bigr)
        \end{aligned} \\
        + & c_3
        \begin{aligned}[t]
            s_{12} (s_{13}-s_{23}) \bigl((l_a-q)^2 - l_a^2 \bigr)
        \end{aligned} \\
        + & c_4
        \begin{aligned}[t]
            (s_{13}^2-s_{23}^2) \bigl((l_a-q)^2-l_a^2 \bigr)
        \end{aligned}
    \end{aligned} $
    & $N_c^2 \tilde f^{a_1 a_2 a_3} $ & 1 \\
    \hline \\
    %%%%%%%%%%%%%%% top 5 information %%%%%%%%%%%%%%%
    \begin{tabular}{cc}    
        \includegraphics[width=0.20\linewidth]{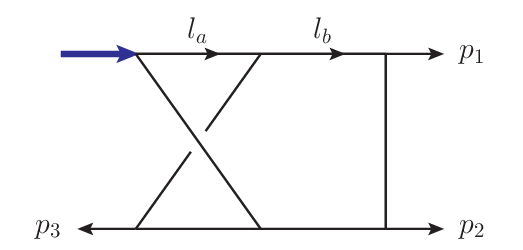}  \\
        (5)  
    \end{tabular} 
    & 
    $\begin{aligned}
          & s_{12} (s_{23}q^2+s_{13}\tau_{a1}-s_{23}\tau_{a2})/2 \\
        + & c_1
        \begin{aligned}[t]
            \bigl( & s_{13} (s_{12}-s_{23}) (-l_b+p_1+p_2)^2 + s_{23} (s_{12}-s_{13}) l_b^2 \\
            +      & s_{12} (s_{13}-s_{23}) (-l_a+l_b+p_3)^2 - s_{12} (s_{13}-s_{23}) (l_a-l_b)^2 \\
            -      & (s_{12}(s_{13}+s_{23})-2s_{13}s_{23}) (l_b-p_1)^2 \bigr)
        \end{aligned} \\
        + & c_2
        \begin{aligned}[t]
            \bigl( & (s_{12}^2-s_{23}^2) (-l_b+p_1+p_2)^2 + (s_{12}^2-s_{13}^2) l_b^2 \\
            +      & (s_{13}^2-s_{23}^2) (-l_a+l_b+p_3)^2 - (s_{13}^2-s_{23}^2) (l_a-l_b)^2 \\
            -      & (2s_{12}^2-s_{13}^2-s_{23}^2) (l_b-p_1)^2\bigr)
        \end{aligned} \\
        - & c_3
        \begin{aligned}[t]
            s_{12} (s_{13}-s_{23}) ((l_a-q)^2 - l_a^2)
        \end{aligned} \\
        - & c_4
        \begin{aligned}[t]
            (s_{13}^2-s_{23}^2) ((l_a-q)^2 - l_a^2)
        \end{aligned}
    \end{aligned} $
    & $N_c^2 \tilde f^{a_1 a_2 a_3} $ & 2 \\
    \hline \\
    %%%%%%%%%%%%%%% top 6 information %%%%%%%%%%%%%%%
    \begin{tabular}{cc}    
        \includegraphics[width=0.20\linewidth]{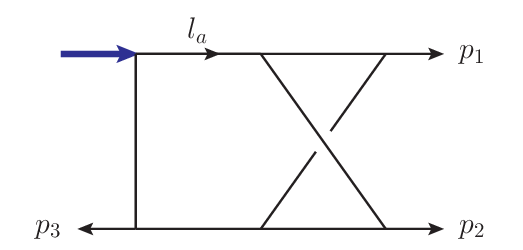}  \\
        (6)  
    \end{tabular} 
    & 
    $\begin{aligned}
          & s_{12} (\tau_{a1}s_{13}-\tau_{a2}s_{23})/2 \\
        - & c_1
        \begin{aligned}[t]
            s_{12} (s_{13}-s_{23}) ((l_a-p_1-p_2)^2-(l_a-q)^2)
        \end{aligned} \\
        - & c_2
        \begin{aligned}[t]
            (s_{13}^2-s_{23}^2) ((l_a-p_1-p_2)^2-(l_a-q)^2)
        \end{aligned} \\
        - & c_3
        \begin{aligned}[t]
            s_{12}(s_{13}-s_{23}) ((l_a-q)^2-l_a^2)
        \end{aligned} \\
        - & c_4
        \begin{aligned}[t]
            (s_{13}^2-s_{23}^2) ((l_a-q)^2-l_a^2)
        \end{aligned}
    \end{aligned} $
    & $0 $ & 2 \\
    \hline
\end{longtable}
%%%%%%%%%%%%%%%%%%% 

The CK-dual numerators contain four free parameters $\{c_1, c_2, c_3, c_4\}$.
As discuss in the paper, they cancel at the integrand level and originate from generalized gauge transformations (GGT) of both Jacobi-induced and operator-induced types.
Below we provide explicit details of these cancellations based on the two-loop example.

The cancellations can be decomposed into eight sub-groups,
which are listed in Table~\ref{tab:fifth_class_ckdual_preserving_GGT}-~\ref{tab:forth_class_ckdual_preserving_GGT}. 
In these 8 tables, $\Gamma_{s,t,u}$ refer to the top-level diagrams in  Jacobi-induced GGTs, while $\Gamma_{a,b}$ mean the top-level diagrams in  operator-induced ones. And $\Gamma$ on the right hand side denotes the one-propagator-reduced diagram, similar to the diagram (c) in Fig.~8 in the paper. 
In the lower part of each table, we collect the terms depending on every appeared free parameter $c_i$, and show that these terms indeed cancel via the mechanism of the GGTs  after summing top-level diagrams.

To be more explicit, consider Table~\ref{tab:fifth_class_ckdual_preserving_GGT}. In this case, we focus on two top-level graphs $\Gamma_a$ and $\Gamma_b$. From the numerator expressions in  Table~\ref{tab:resphi2-2loop}, we can read the terms that depend on parameter $c_3$ and are also proportional to the propagators $l_A$ and $l_B$, respectively:
\begin{equation}
N_{\Gamma_a} \big|_{c_3 l_A^2} = -s_{12}(s_{13}-s_{23}) \,, \qquad N_{\Gamma_b} \big|_{c_3 l_B^2} = +s_{12}(s_{13}-s_{23}) \,.
\end{equation}
After shrinking the propagators $l_A$ and $l_B$, both graphs reduce to the one-propagator-reduced diagram $\Gamma$. Together with their color factors, these two terms indeed cancel with each other:
\begin{equation}
 C_a I_a\big[-s_{12}(s_{13}-s_{23})l_A^2 \big]  + C_b I_b \big[ s_{12}(s_{13}-s_{23})   l_B^2 \big]  =  -s_{12}(s_{13}-s_{23}) (C_a - C_b)I_\Gamma[1] = 0 \,,
\end{equation}
due to the operator-induced color relation $C_a-C_b=0$, as also presented in Table~\ref{tab:fifth_class_ckdual_preserving_GGT}.

We stress that the non-trivial point is that these deformations preserves all dual Jacobi relations, and to achieve this, it is necessary to involve operator-induced GGTs.

%%%%%%%%%%%%%%% class of GGTs %%%%%%%%%%%%%%%
%%%%%%%%%%%%%%%%%%% 
\begin{longtable}{|cc|c|}
    \caption{Operator-induced GGTs (1)}
    \label{tab:fifth_class_ckdual_preserving_GGT}
    \endfirsthead
    \endhead
    \hline \\
    %%%%%%%%%%%%%%% graph information %%%%%%%%%%%%%%%
    \begin{tabular}{c}
        \includegraphics[width=0.20\linewidth]{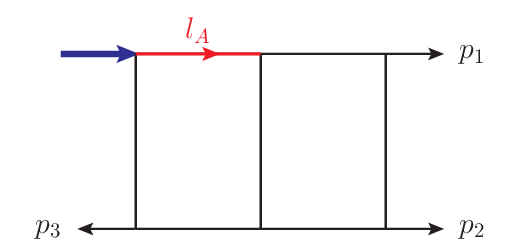}  \\
        $\Gamma_a$
        %$\Gamma_a = \Gamma_{4,123} $
    \end{tabular} &
    \begin{tabular}{c}
        \includegraphics[width=0.20\linewidth]{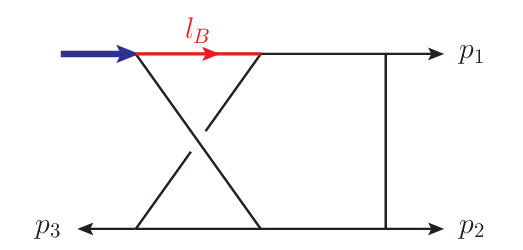}  \\
        $\Gamma_b$
        %$\Gamma_b = \Gamma_{5,123} $
    \end{tabular} &
    \begin{tabular}{c}
        \includegraphics[width=0.20\linewidth]{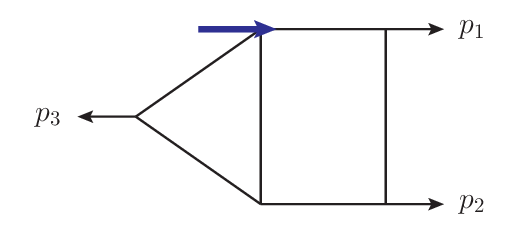}  \\
        $\Gamma $ \\
        %$\Gamma = \Gamma_{6:5,123} $ \\
        $C_a - C_b = 0 $
    \end{tabular} \\ 
    \hline \\
    %%%%%%%%%%%%%%% GGT information %%%%%%%%%%%%%%%
    $c_3 $ &
    \multicolumn{2}{|c|}{
    $\begin{aligned}
          & -s_{12}(s_{13}-s_{23}) \bigl(C_a I_a[l_A^2] - C_b I_b[l_B^2] \bigr) \\
        = & -s_{12}(s_{13}-s_{23}) (C_a - C_b)I_\Gamma[1] = 0
    \end{aligned} $
    } \\
    \hline \\
    $c_4 $ &
    \multicolumn{2}{|c|}{
    $\begin{aligned}
          & -(s_{13}^2-s_{23}^2) \bigl(C_a I_a[l_A^2] - C_b I_b[l_B^2] \bigr) \\
        = & -(s_{13}^2-s_{23}^2) (C_a - C_b)I_\Gamma[1] = 0
    \end{aligned} $
    } \\
    \hline
\end{longtable}
%%%%%%%%%%%%%%%%%%% 

%%%%%%%%%%%%%%% class of GGTs %%%%%%%%%%%%%%%
%%%%%%%%%%%%%%%%%%% 
\begin{longtable}{|cc|c|}
    \caption{Operator-induced GGTs (2)}
    \label{tab:sixth_class_ckdual_preserving_GGT}
    \endfirsthead
    \endhead
    \hline \\
    %%%%%%%%%%%%%%% graph information %%%%%%%%%%%%%%%
    \begin{tabular}{c}
        \includegraphics[width=0.20\linewidth]{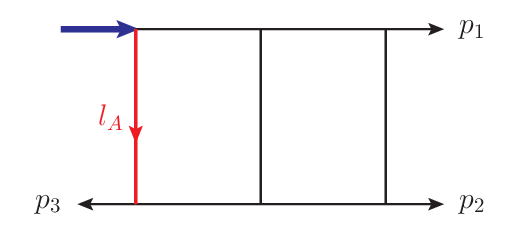}  \\
        $\Gamma_a $
        %$\Gamma_a = \Gamma_{4,123} $
    \end{tabular} &
    \begin{tabular}{c}
        \includegraphics[width=0.20\linewidth]{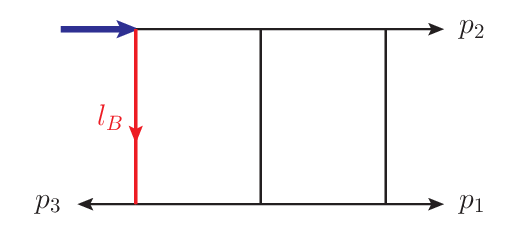}  \\
        $\Gamma_b$
        %$\Gamma_b = \Gamma_{4,213} $
    \end{tabular} &
    \begin{tabular}{c}
        \includegraphics[width=0.20\linewidth]{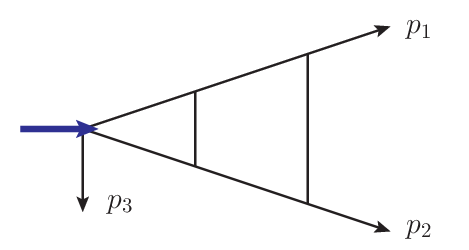}  \\
        $\Gamma $ \\
        %$\Gamma = \Gamma_{6:6,123} $ \\
        $C_a - C_b = 0 $
    \end{tabular} \\ 
    \hline \\
    %%%%%%%%%%%%%%% GGT information %%%%%%%%%%%%%%%
    $c_1 $ &
    \multicolumn{2}{|c|}{
    $\begin{aligned}
          & -s_{12}(s_{13}-s_{23}) \bigl(C_a I_a[l_A^2] - C_b I_b[l_B^2] \bigr) \\
        = & -s_{12}(s_{13}-s_{23}) (C_a - C_b)I_\Gamma[1] = 0
    \end{aligned} $
    } \\
    \hline \\
    $c_2 $ &
    \multicolumn{2}{|c|}{
    $\begin{aligned}
          & -(s_{13}^2-s_{23}^2) \bigl(C_a I_a[l_A^2] - C_b I_b[l_B^2] \bigr) \\
        = & -(s_{13}^2-s_{23}^2) (C_a - C_b)I_\Gamma[1] = 0
    \end{aligned} $
    } \\
    \hline \\
    $c_3 $ &
    \multicolumn{2}{|c|}{
    $\begin{aligned}
          & s_{12}(s_{13}-s_{23}) \bigl(C_a I_a[l_A^2] - C_b I_b[l_B^2] \bigr) \\
        = & s_{12}(s_{13}-s_{23}) (C_a - C_b)I_\Gamma[1] = 0
    \end{aligned} $
    } \\
    \hline \\
    $c_4 $ &
    \multicolumn{2}{|c|}{
    $\begin{aligned}
          & (s_{13}^2-s_{23}^2) \bigl(C_a I_a[l_A^2] - C_b I_b[l_B^2] \bigr) \\
        = & (s_{13}^2-s_{23}^2) (C_a - C_b)I_\Gamma[1] = 0
    \end{aligned} $
    } \\
    \hline
\end{longtable}
%%%%%%%%%%%%%%%%%%% 
\newpage

%%%%%%%%%%%%%%% class of GGTs %%%%%%%%%%%%%%%
%%%%%%%%%%%%%%%%%%% 
\begin{longtable}{|cc|c|}
    \caption{Operator-induced GGTs (3)}
    \label{tab:seventh_class_ckdual_preserving_GGT}
    \endfirsthead
    \endhead
    \hline \\
    %%%%%%%%%%%%%%% graph information %%%%%%%%%%%%%%%
    \begin{tabular}{c}
        \includegraphics[width=0.20\linewidth]{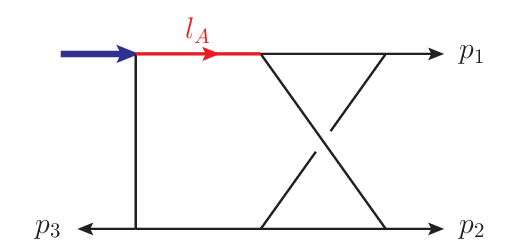}  \\
        $\Gamma_a $
        %$\Gamma_a = \Gamma_{6,123} $
    \end{tabular} &
    \begin{tabular}{c}
        \includegraphics[width=0.20\linewidth]{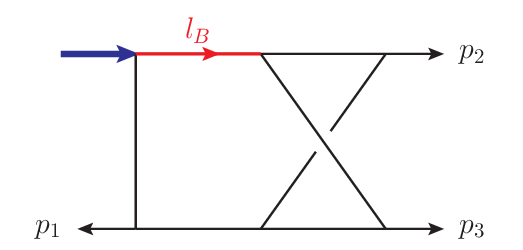}  \\
         $\Gamma_b$
        %$\Gamma_b = \Gamma_{6,231} $
    \end{tabular} &
    \begin{tabular}{c}
        \includegraphics[width=0.20\linewidth]{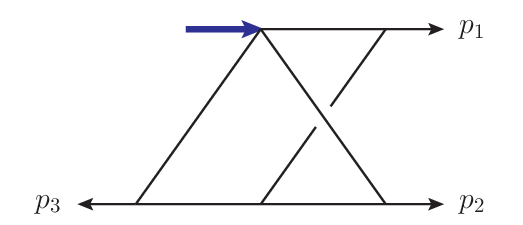}  \\
         $\Gamma $ \\
        %$\Gamma = \Gamma_{6:7,123} $ \\
        $C_a - C_b = 0 $
    \end{tabular} \\ 
    \hline \\
    %%%%%%%%%%%%%%% GGT information %%%%%%%%%%%%%%%
    $c_3 $ &
    \multicolumn{2}{|c|}{
    $\begin{aligned}
          & s_{12}s_{13} \bigl(C_a I_a[l_A^2] - C_b I_b[l_B^2] \bigr) \\
        = & s_{12}s_{13} (C_a - C_b)I_\Gamma[1] = 0
    \end{aligned} $
    } \\
    \hline \\
    $c_4 $ &
    \multicolumn{2}{|c|}{
    $\begin{aligned}
          & -s_{23}^2 \bigl(C_a I_a[l_A^2] - C_b I_b[l_B^2] \bigr) \\
        = & -s_{23}^2 (C_a - C_b)I_\Gamma[1] = 0
    \end{aligned} $
    } \\
    \hline
\end{longtable}
%%%%%%%%%%%%%%%%%%% 

%%%%%%%%%%%%%%% class of GGTs %%%%%%%%%%%%%%%
%%%%%%%%%%%%%%%%%%% 
\begin{longtable}{|cc|c|}
    \caption{Operator-induced GGTs (4)}
    \label{tab:eighth_class_ckdual_preserving_GGT}
    \endfirsthead
    \endhead
    \hline \\
    %%%%%%%%%%%%%%% graph information %%%%%%%%%%%%%%%
    \begin{tabular}{c}
        \includegraphics[width=0.20\linewidth]{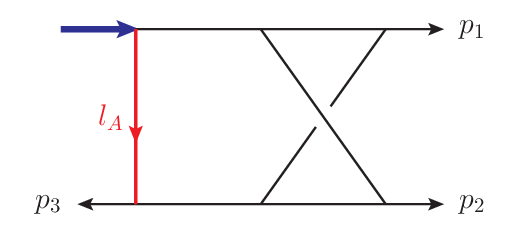}  \\
        $\Gamma_a $   
       %$\Gamma_a = \Gamma_{4,123} $
    \end{tabular} &
    \begin{tabular}{c}
        \includegraphics[width=0.20\linewidth]{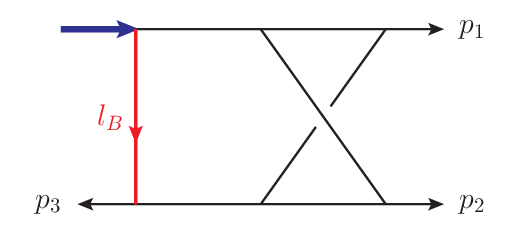}  \\
        $\Gamma_b$
       % $\Gamma_b = \Gamma_{4,123} $
    \end{tabular} &
    \begin{tabular}{c}
        \includegraphics[width=0.20\linewidth]{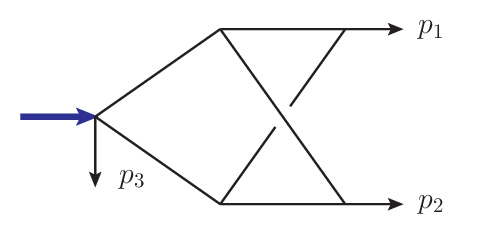}  \\
         $\Gamma $ \\
        %$\Gamma = \Gamma_{6:8,123} $ \\
        $C_a - C_b = 0 $
    \end{tabular} \\ 
    \hline \\
    %%%%%%%%%%%%%%% GGT information %%%%%%%%%%%%%%%
    $c_1 $ &
    \multicolumn{2}{|c|}{
    $\begin{aligned}
          & 1/2 s_{12}(s_{13}-s_{23}) \bigl(C_a I_a[l_A^2] - C_b I_b[l_B^2] \bigr) \\
        = & 1/2 s_{12}(s_{13}-s_{23}) (C_a - C_b)I_\Gamma[1] = 0
    \end{aligned} $
    } \\
    \hline \\
    $c_2 $ &
    \multicolumn{2}{|c|}{
    $\begin{aligned}
          & 1/2 (s_{13}^2-s_{23}^2) \bigl(C_a I_a[l_A^2] - C_b I_b[l_B^2] \bigr) \\
        = & 1/2 (s_{13}^2-s_{23}^2) (C_a - C_b)I_\Gamma[1] = 0
    \end{aligned} $
    } \\
    \hline \\
    $c_3 $ &
    \multicolumn{2}{|c|}{
    $\begin{aligned}
          & -1/2 s_{12}(s_{13}-s_{23}) \bigl(C_a I_a[l_A^2] - C_b I_b[l_B^2] \bigr) \\
        = & -1/2 s_{12}(s_{13}-s_{23}) (C_a - C_b)I_\Gamma[1] = 0
    \end{aligned} $
    } \\
    \hline \\
    $c_4 $ &
    \multicolumn{2}{|c|}{
    $\begin{aligned}
          & -1/2 (s_{13}^2-s_{23}^2) \bigl(C_a I_a[l_A^2] - C_b I_b[l_B^2] \bigr) \\
        = & -1/2 (s_{13}^2-s_{23}^2) (C_a - C_b)I_\Gamma[1] = 0
    \end{aligned} $
    } \\
    \hline
\end{longtable}
%%%%%%%%%%%%%%%%%%% 

%%%%%%%%%%%%%%% class of GGTs %%%%%%%%%%%%%%%
%%%%%%%%%%%%%%%%%%% 
\begin{longtable}{|ccc|c|}
    \caption{Jacobi-induced GGTs (1)}
    \label{tab:first_class_ckdual_preserving_GGT}
    \endfirsthead
    \endhead
    \hline \\
    %%%%%%%%%%%%%%% graph information %%%%%%%%%%%%%%%
    \begin{tabular}{c}
        \includegraphics[width=0.20\linewidth]{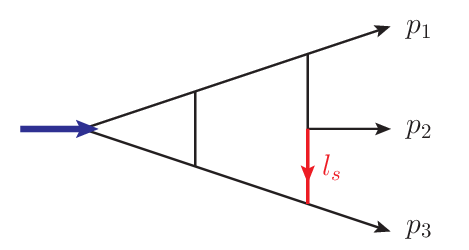}  \\
        $\Gamma_s $
        %$\Gamma_s = \Gamma_{1,123} $
    \end{tabular} &
    \begin{tabular}{c}
        \includegraphics[width=0.20\linewidth]{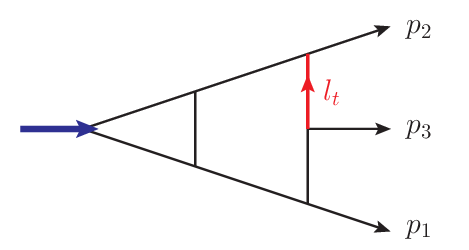}  \\
        $\Gamma_t $
        %$\Gamma_t = \Gamma_{1,231} $
    \end{tabular} &
    \begin{tabular}{c}
        \includegraphics[width=0.20\linewidth]{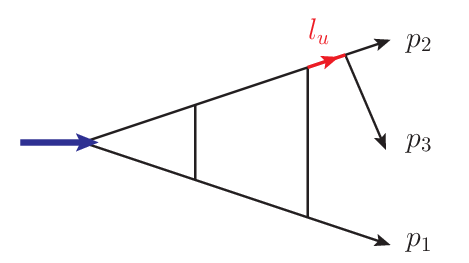}  \\
        $\Gamma_u $
        %$\Gamma_u = \Gamma_{2,231} $
    \end{tabular} &
    \begin{tabular}{c}
        \includegraphics[width=0.20\linewidth]{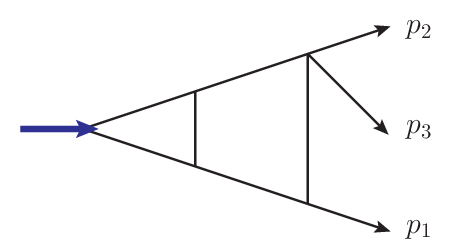}  \\
        $\Gamma$ \\
        %$\Gamma = \Gamma_{6:1,231} $ \\
        $C_s + C_t - C_u = 0 $
    \end{tabular} \\ 
    \hline \\
    %%%%%%%%%%%%%%% GGT information %%%%%%%%%%%%%%%
    $c_1 $ &
    \multicolumn{3}{|c|}{
    $\begin{aligned}
          & -\bigl(s_{23}(s_{12}+s_{13})-2s_{12}s_{13} \bigr) \bigl(C_s I_s[l_s^2] + C_t I_t[l_t^2] - C_u I_u[l_u^2] \bigr) \\
        = & -\bigl(s_{23}(s_{12}+s_{13})-2s_{12}s_{13} \bigr) (C_s + C_t - C_u)I_\Gamma[1] = 0
    \end{aligned} $
    } \\
    \hline \\
    $c_2 $ &
    \multicolumn{3}{|c|}{
    $\begin{aligned}
          & -\bigl(2s_{23}^2 - s_{12}^2 - s_{13}^2 \bigr) \bigl(C_s I_s[l_s^2] + C_t I_t[l_t^2] - C_u I_u[l_u^2] \bigr) \\
        = & -\bigl(2s_{23}^2 - s_{12}^2 - s_{13}^2 \bigr) (C_s + C_t - C_u)I_\Gamma[1] = 0
    \end{aligned} $
    } \\
    \hline
\end{longtable}
%%%%%%%%%%%%%%%%%%% 

\newpage
%%%%%%%%%%%%%%% class of GGTs %%%%%%%%%%%%%%%
%%%%%%%%%%%%%%%%%%% 
\begin{longtable}{|ccc|c|}
    \caption{Jacobi-induced GGTs (2)}
    \label{tab:second_class_ckdual_preserving_GGT}
    \endfirsthead
    \endhead
    \hline \\
    %%%%%%%%%%%%%%% graph information %%%%%%%%%%%%%%%
    \begin{tabular}{c}
        \includegraphics[width=0.20\linewidth]{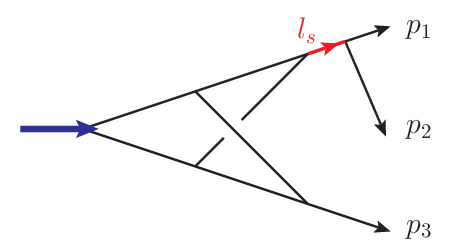}  \\
         $\Gamma_s $
        %$\Gamma_s = \Gamma_{3,123} $
    \end{tabular} &
    \begin{tabular}{c}
        \includegraphics[width=0.20\linewidth]{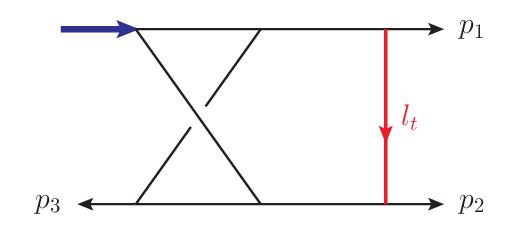}  \\
        $\Gamma_t $
        %$\Gamma_t = \Gamma_{5,123} $
    \end{tabular} &
    \begin{tabular}{c}
        \includegraphics[width=0.20\linewidth]{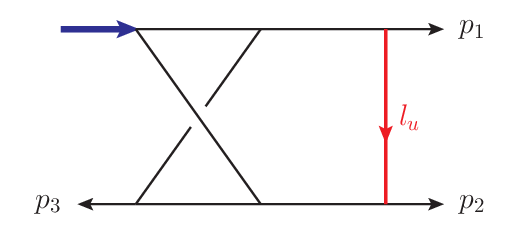}  \\
         $\Gamma_u $
        %$\Gamma_u = \Gamma_{5,123} $
    \end{tabular} &
    \begin{tabular}{c}
        \includegraphics[width=0.20\linewidth]{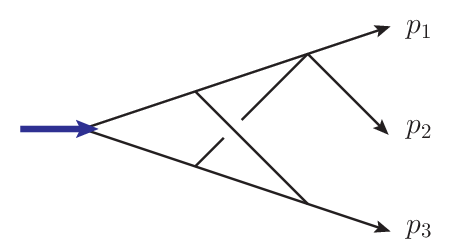}  \\
        $\Gamma $ \\
        %$\Gamma = \Gamma_{6:2,123} $ \\
        $C_s - C_t - C_u = 0 $
    \end{tabular} \\ 
    \hline \\
    %%%%%%%%%%%%%%% GGT information %%%%%%%%%%%%%%%
    $c_1 $ &
    \multicolumn{3}{|c|}{
    $\begin{aligned}
          & 1/2 \bigl(s_{12}(s_{13}+s_{23})-2s_{13}s_{23} \bigr) \bigl(C_s I_s[l_s^2] - C_t I_t[l_t^2] - C_u I_u[l_u^2] \bigr) \\
        = & 1/2 \bigl(s_{12}(s_{13}+s_{23})-2s_{13}s_{23} \bigr) (C_s - C_t - C_u)I_\Gamma[1] = 0
    \end{aligned} $
    } \\
    \hline \\
    $c_2 $ &
    \multicolumn{3}{|c|}{
    $\begin{aligned}
          & 1/2 \bigl(2s_{12}^2 - s_{13}^2 - s_{23}^2 \bigr) \bigl(C_s I_s[l_s^2] - C_t I_t[l_t^2] - C_u I_u[l_u^2] \bigr) \\
        = & 1/2 \bigl(2s_{12}^2 - s_{13}^2 - s_{23}^2 \bigr) (C_s - C_t - C_u)I_\Gamma[1] = 0
    \end{aligned} $
    } \\
    \hline
\end{longtable}
%%%%%%%%%%%%%%%%%%% 

%%%%%%%%%%%%%%% class of GGTs %%%%%%%%%%%%%%%
%%%%%%%%%%%%%%%%%%% 
\begin{longtable}{|ccc|c|}
    \caption{Jacobi-induced GGTs (3)}
    \label{tab:third_class_ckdual_preserving_GGT}
    \endfirsthead
    \endhead
    \hline \\
    %%%%%%%%%%%%%%% graph information %%%%%%%%%%%%%%%
    \begin{tabular}{c}
        \includegraphics[width=0.20\linewidth]{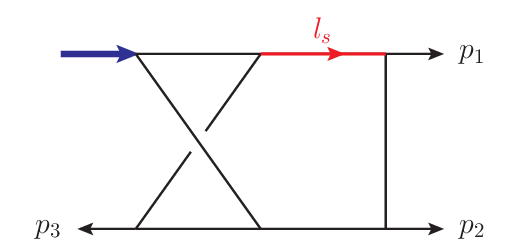}  \\
        $\Gamma_s $
        %$\Gamma_s = \Gamma_{5,123} $
    \end{tabular} &
    \begin{tabular}{c}
        \includegraphics[width=0.20\linewidth]{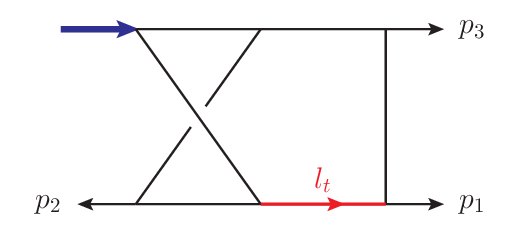}  \\
         $\Gamma_t$
        %$\Gamma_t = \Gamma_{5,312} $
    \end{tabular} &
    \begin{tabular}{c}
        \includegraphics[width=0.20\linewidth]{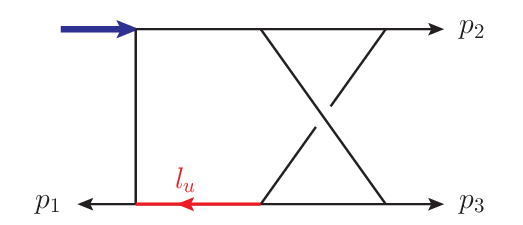}  \\
        $\Gamma_u $
        %$\Gamma_u = \Gamma_{6,231} $
    \end{tabular} &
    \begin{tabular}{c}
        \includegraphics[width=0.20\linewidth]{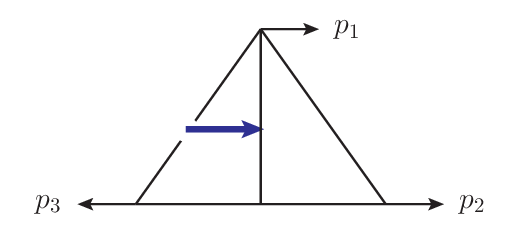}  \\
        $\Gamma $ \\
        %$\Gamma = \Gamma_{6:3,123} $ \\
        $C_s - C_t - C_u = 0 $
    \end{tabular} \\ 
    \hline \\
    %%%%%%%%%%%%%%% GGT information %%%%%%%%%%%%%%%
    $c_1 $ &
    \multicolumn{3}{|c|}{
    $\begin{aligned}
          & s_{23}(s_{12}-s_{13}) \bigl(C_s I_s[l_s^2] - C_t I_t[l_t^2] - C_u I_u[l_u^2] \bigr) \\
        = & s_{23}(s_{12}-s_{13}) (C_s - C_t - C_u)I_\Gamma[1] = 0
    \end{aligned} $
    } \\
    \hline \\
    $c_2 $ &
    \multicolumn{3}{|c|}{
    $\begin{aligned}
          & (s_{12}^2-s_{13}^2) \bigl(C_s I_s[l_s^2] - C_t I_t[l_t^2] - C_u I_u[l_u^2] \bigr) \\
        = & (s_{12}^2-s_{13}^2) (C_s - C_t - C_u)I_\Gamma[1] = 0
    \end{aligned} $
    } \\
    \hline
\end{longtable}
%%%%%%%%%%%%%%%%%%% 

%%%%%%%%%%%%%%% class of GGTs %%%%%%%%%%%%%%%
%%%%%%%%%%%%%%%%%%% 
\begin{longtable}{|ccc|c|}
    \caption{Jacobi-induced GGTs (4)}
    \label{tab:forth_class_ckdual_preserving_GGT}
    \endfirsthead
    \endhead
    \hline \\
    %%%%%%%%%%%%%%% graph information %%%%%%%%%%%%%%%
    \begin{tabular}{c}
        \includegraphics[height=0.09\linewidth]{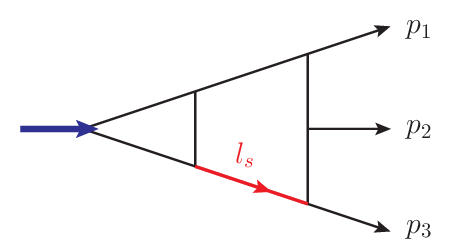}  \\
        $\Gamma_s $
        %$\Gamma_s = \Gamma_{1,123} $
    \end{tabular} &
    \begin{tabular}{c}
        \includegraphics[width=0.20\linewidth]{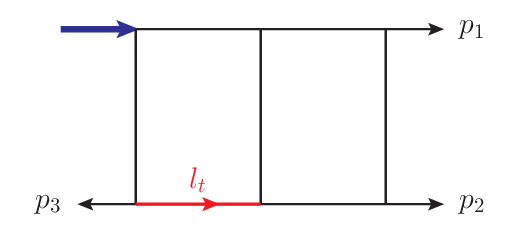}  \\
         $\Gamma_t$
        %$\Gamma_t = \Gamma_{4,123} $
    \end{tabular} &
    \begin{tabular}{c}
        \includegraphics[width=0.20\linewidth]{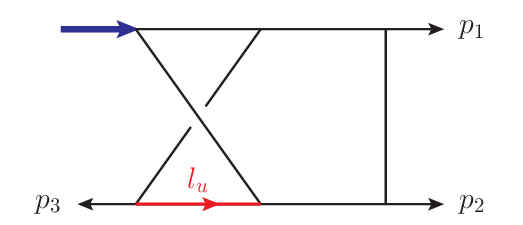}  \\
        $\Gamma_u$
        %$\Gamma_u = \Gamma_{5,123} $
    \end{tabular} &
    \begin{tabular}{c}
        \includegraphics[width=0.20\linewidth]{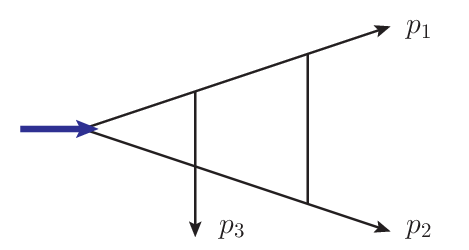}  \\
        $\Gamma$ \\
        %$\Gamma = \Gamma_{6:4,123} $ \\
        $C_s - C_t - C_u = 0 $
    \end{tabular} \\ 
    \hline \\
    %%%%%%%%%%%%%%% GGT information %%%%%%%%%%%%%%%
    $c_1 $ &
    \multicolumn{3}{|c|}{
    $\begin{aligned}
          & -s_{12}(s_{13}-s_{23}) \bigl(C_s I_s[l_s^2] - C_t I_t[l_t^2] - C_u I_u(l_u^2) \bigr) \\
        = & -s_{12}(s_{13}-s_{23}) (C_s - C_t - C_u)I_\Gamma[1] = 0
    \end{aligned} $
    } \\
    \hline \\
    $c_2 $ &
    \multicolumn{3}{|c|}{
    $\begin{aligned}
          & -(s_{13}^2-s_{23}^2) \bigl(C_s I_s[l_s^2] - C_t I_t[l_t^2] - C_u I_u(l_u^2) \bigr) \\
        = & -(s_{13}^2-s_{23}^2) (C_s - C_t - C_u)I_\Gamma[1] = 0
    \end{aligned} $
    } \\
    \hline
\end{longtable}
%%%%%%%%%%%%%%%%%%% 

\end{document}